\documentclass[aps,pra,twocolumn,showpacs,groupedaddress]{revtex4}
\usepackage{graphicx,graphics,times}
\begin{document}
\title{ New Approach to Quantum Key Distribution Via Quantum Encryption}
\author{A. Fahmi }
\email{ fahmi@theory.ipm.ac.ir}


\affiliation{ Institute for Studies in Theoretical Physics and
Mathematics (IPM) P. O. Box 19395-5531, Tehran, Iran}

\begin{abstract}
In this Paper, we investigate the security of Zhang, Li and Guo quantum key distribution via quantum encryption protocol
[$\text{Phys. Rev. A} \textbf{64}, 24302 (2001)$] and show that it is not secure against some of Eve's attacks and with
the probability one half she gets all of keys without being detected by the two parties. The main defect in this protocol is that
there is an attack strategy by which Eve can change the previously shared Bell state between Alice and Bob to two Bell
states among herself and Alice and Bob. Hence, we show that with probability $1/d$ its generalization to $d$-dimension
systems is not secure and show that its extension to the case of
more partners based on the reusable GHZ states is not secure and with probability one half Eve gets all of keys
without being detected by the two parties. In what follows, we show how in going
to higher dimensions those protocols can be repaired.
\end{abstract}
 \maketitle
\section{Introduction}
In a cryptography protocol, we discus the possibility that two remote parties, conventionally called Alice and Bob,
exchange a secret random key to implement a secure encryption-decryption algorithm, without meeting each other.
Once the sharing is achieved, the two parties can secretly exchange a message over the public channel by encrypting
them with a key with a length equal to the message. In the key distribution with classical transmission lines,
an eavesdropper can freely sneak into the transmissions and monitor the information. Thus the role of cryptography
is to provide some mathematical procedure that makes it computationally difficult for the eavesdropper to reproduce
the key from the information sent through the transmission lines. However, no existing classical cryptosystems
has proven to present sufficient difficulty to an eavesdropper. In particular, it was shown that some of them
can be broken in principle by quantum computation \cite{Shor}.

On the other hand, quantum mechanics (QM) and its fundamental concepts were previously restricted to clarify and
describe very special issues of physics. Recently, these concepts have been extensively used in such concrete problems
as the distribution of secure keys between two parties. In 1984, Bennett and Brassard \cite{BB} proposed a
way of distributing the keys in a physically secure way, by using quantum physics. Their protocol bears the acronym BB84,
and was the first protocol of quantum cryptography, but we shall use the more precise name of quantum key distribution
(QKD).

In 1991, Ekert \cite{Ek} proposed a QKD protocol that used entangled particles, and he stated that the violation
of Bell's inequality might be the physical principle that ensures security. In the quantum mechanical approach to key
distribution, security is based on the laws of QM and not on the unproven complexity of a mathematical problem, as was
the case with the classical cryptography. In these protocols eavesdropping by an Eve has been considered, and it has been
shown that they are secure against various types of Eve's attacks (for a comprehensive review see \cite{Gis1}).

After the earlier QKD protocols, various extensions were proposed, for example: a key distribution protocol in which
Alice distributes the key among $N$ different Bobs, in such a way that only with the participation of all $m$ Bobs
$(m \leq N)$ the keys could be retrieved \cite{Gis2}. Continuous variable key distributions \cite{Gro} are
among these generalizations. QKD protocols have been demonstrated over the past fifteen years through an ingenious
series of experiments by several research groups. Many encouraging experiments have demonstrated QKD, some spanning more
than a hundred kilometers through optical fibers \cite{Mu}. Hence, the possibility of a secure key distribution in
the presence of noisy channels and decoherence has been considered \cite{Brass}.

Recently, a new approach to QKD was proposed by Zhang, Li and Guo (ZLG) \cite{ZLG}, in which they proposed a QKD
protocol that uses a quantum key to encode and decode the classical information by considering a previously shared
reusable Bell state which acts as the quantum key. In this paper, we show that their protocol is insecure against
Eve's attack and with a probability of $1/2$, Eve can be informed of Alice and Bob keys without being detected
by them. Hence, we show that their extensions to higher dimension \cite{KBB} and quantum secret sharing based on
the reusable GHZ states \cite{BK}, are not secure.

This paper is organized as follows. In Sec. II we briefly review some properties of QKD protocols in which data exchange
between Alice and Bob is done, using orthogonal states. In Sec. III we review the original Zhang \emph{et al.} protocol
\cite{ZLG}. In Sec. IV we show that this the protocol is not secure against some of Eve's attacks. In Sec. V we show that
ZLG extension to higher dimensions, suggested by Karimipour, Bahraminasab and Bagherinezhad (KBB) \cite{KBB}
is not secure. In Sec. VI we present some modified protocols that are not only secure against Eve's attacks, but also our
protocol has simple QKD steps and any eavesdropping by Eve can be detected by Alice and Bob. Hence, we show that ZLG
protocol can be repaired, if we go to higher dimensional shared EPR states. In Sec. VII we show that quantum secret
sharing based on the reusable GHZ state among three parties, suggested by Bagherinezhad and
Karimipour (Bk) \cite{BK}, is not secure and we repair this protocol. Finally, in Sec. VIII we summarize our results.

\section{Quantum Cryptography Based on Orthogonal States}
Security of quantum cryptosystems is guaranteed by the two fundamental laws of QM, (i) no-cloning theorem of
non-orthogonal states in quantum mechanics, (ii) any measurement on a system can cause a perturbation on it
(wave function reduction). On the other hand, in addition to the above basic principle, in a realistic QKD protocol, we
mention some conditions that should be satisfied by any QKD protocol. As suggested by M. Nielsen \cite{Niel}, to get the best
possible cryptosystems, we need to have the following properties:

I) The theoretical basis must be entirely public;

II) Actual implementation should be widely and publicly available;

III) There should be extended periods of testing real implementations.

In addition to the above conditions, any QKD protocol must have the following property,

IV) If Alice and Bob detect the presence of Eve on the distributed keys, they should discard those keys and must apply
their protocol to other states, and in this new use of the protocol they must detect the existence of Eve again.

Although, systems obeying these properties don't necessarily have security, they have preconditions for being secure.
A simple scheme was proposed by Bennett \cite{BB}, which uses only two nonorthogonal states. In this method,
nonorthogonal encoding of the bit of information inevitably leads to the waste of a portion of photons.

In contrast to these schemes that use nonorthogonal states, some deterministic quantum cryptography schemes, based
on orthogonal states, have been proposed \cite{GV, KI, Cab}. Goldenberg and Vaidman (GV) \cite{GV} and consequently,
Koashi and Imoto (KI) \cite{KI} have presented a quantum cryptography based on orthogonal states. The basic technique is to
split the transfer of one bit of information into two steps, ensuring that only a fraction of the bit of information
is transmitted at a time. Then the no-cloning theorem of orthogonal states \cite{Mor} guarantee
its security. Cabello presented a protocol \cite{Cab} based on GV and KI protocols and on the idea of using a larger
alphabet that saturates the capacity of the quantum channel. He has introduced a key distribution protocol which works
in the Holevo limit \cite{Cab}. After this paper, some people attempted to get Holevo limit by using another approach
\cite{Deg,Ma,Lon}. Recently, some people \cite{GLSLG, HL} have used non-local variables, firstly introduced by Bennett
\emph{et al.} \cite{Ben}, to propose protocols for QKD. These state, which are the product of two states cannot be
distinguished through the use of local operations and (unlimited) classical two-way communications (LOCC).

Hence, Zhang \emph{et al.} \cite{ZLG} presented another QKD scheme that uses a quantum key to encode and decode the
classical information, and the previously shared reusable EPR state acts as the quantum key. They called it channel
encrypting QKD. This protocol can be used for quantum secure direct communication, and was proposed and pursued by
some groups \cite{QSDC}.

One of the best properties of QKD with orthogonal states is that we don't use any classical communication
(although, earlier versions didn't have this property), because of the lack of classical communications, Eve has
minimum information about the secret keys that are shared by Alice and Bob. Hence in this protocol, we make use of the
total dimension of Hilbert space. Thus, the security of protocol can be enhanced. In what follows, we restrict ourself
to protocols haveing no decoherence effects and in which all of the parties and eavesdropping
operations are exact and the environment doesn't interact with our system.

\section{ZLG Protocol}
In the Zhang \emph{et al.} protocol \cite{ZLG}, Alice and Bob have previously shared some quantity of the Bell pairs, serving
as the quantum key
\begin{eqnarray}
|\Phi^{+}\rangle=\frac{1}{\sqrt{2}}[|00\rangle_{AB}+|11\rangle_{AB}]
\end{eqnarray}
When the process begins, the two parties rotate their particle's state by angle $\theta$, respectively. The rotation
can be described by
\begin{eqnarray}
R(\theta)=\left(%
\begin{array}{cc}
\cos\theta & \sin\theta \\
-\sin\theta & \cos\theta\\
\end{array}%
\right)
\end{eqnarray}
The state $|\Phi^{+}\rangle$ does not change under bilateral operation of $R(\theta)$. The purpose of this operation
is to prevent the other parties from eavesdropping. Then Alice selects a value of a bit ($0$ or $1$) and prepares a
carrier particle $\gamma$ in the corresponding state $|\psi\rangle$ ($|0\rangle$ or $|1\rangle$), randomly.
The classical bit and the state $|\psi\rangle$ are only known by Alice herself. Alice uses the particle $A$ of the
entangled pairs and $\gamma$ in state $|\psi\rangle$ to do a controlled-NOT (\textsc{cnot}) operation ($A$ is the
controller and $\gamma$ is the target) and the three particles will be in a GHZ state
\begin{eqnarray}\label{3}
|\Psi\rangle=\frac{1}{\sqrt{2}}[|000\rangle_{AB\gamma}+|111\rangle_{AB\gamma}],
\hspace{.5cm} when \hspace{.5cm}|\psi\rangle=|0\rangle
\end{eqnarray}
or
\begin{eqnarray}\label{4}
|\Psi\rangle=\frac{1}{\sqrt{2}}[|001\rangle_{AB\gamma}+|110\rangle_{AB\gamma}],
\hspace{.5cm} when \hspace{.5cm}|\psi\rangle=|1\rangle
\end{eqnarray}
Then, she sends $\gamma$ to Bob. Bob uses his corresponding particle $B$ to do a \textsc{cnot} operation on $\gamma$ again
($B$ is the controller and $\gamma$ is the target). Now the key particles $A$ and $B$ and the carrier particle
$\gamma$ are in the same state as the initial state:

\begin{eqnarray}
|\Psi'\rangle=|\Phi^{+}\rangle\otimes|\psi\rangle
\end{eqnarray}
Bob measures $\gamma$ and gets the classical bit corresponding to the state $|\psi\rangle$. Zhang
\emph{et al.} considered some of Eve's eavesdropping strategies, for example, they showed that Eve can intercepts
the particle $\gamma$ that Alice sends to Bob and uses her own particle (in the states $|0\rangle$ or $|1\rangle$)
to do a \textsc{cnot} operation (her own particle is the target and $\gamma$ is the controller). Then, Eve sends
$\gamma$ to Bob. After Bob's decoding operation, Eve's particle is entangled with the key. It seems that Eve can use
her particle to decode Alice's particle as Bob does. To detect this eavesdropping strategy, they suggested that Alice
and Bob should do a bilateral rotation $R(\theta)$ on the key (EPR pairs in the state $|\Phi^{+}\rangle$) before Alice
does the \textsc{cnot} operation for second $\gamma$ particle. Zhang \emph{et al.} showed that the error rate of the
bit that Alice sends to Bob will be $2\cos^{2}\theta \sin^{2}\theta$. So if $\theta=\frac{\pi}{4} $, the error rate
caused by eavesdropping will reach $\frac{1}{2}$. Then they concluded that communication parties can select
$\theta=\frac{\pi}{4}$ as the bilateral rotation angle in every round and Eve cannot get any useful information of
the bit string they transmit. Then, they considered a more realistic quantum operation and transition
through noisy channels and showed that their protocol was robust.

\section{EVE'S ATTACK}
In this section, we would like to show that ZLG protocol is not secure against the use of a special strategy for
eavesdropping.

\begin{figure}
\centering
\includegraphics[height=5cm,width=8cm]{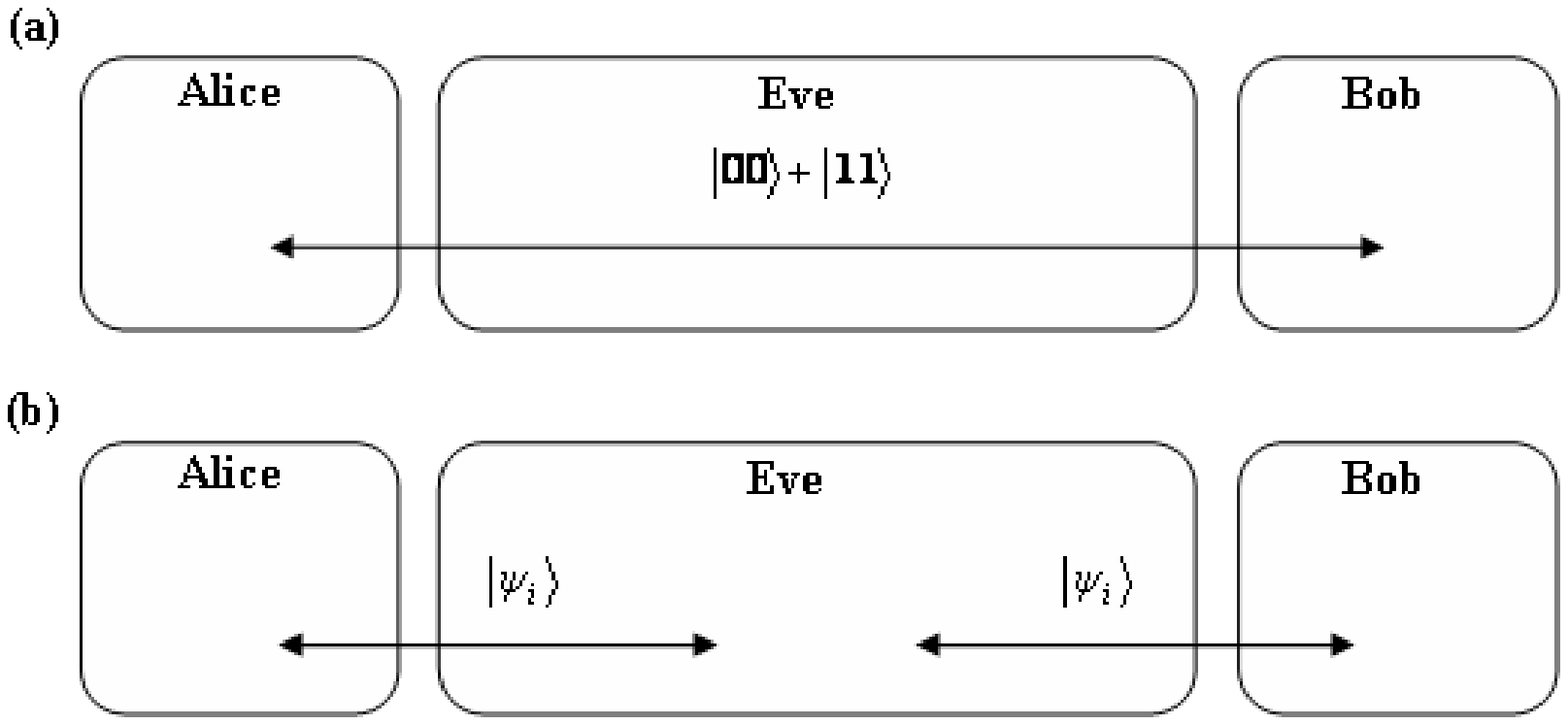}
\caption{A F-type attack on ZLG protocol for quantum encryption of
keys. (a) With the choice of appropriate bases, Eve can change the
previously shared Bell pairs $|\Phi^{+}\rangle_{AB}$ between Alice
and Bob to (b) two similar Bell states between her and Alice and
Bob (b).} \label{Bell}
\end{figure}
The eavesdropping strategy is illustrated in Fig.(\ref{Bell}). According to Fig.(\ref{Bell}), if Eve can transform
previously shared entangled state $|\Phi^{+}\rangle_{AB}$ between Alice and Bob to
$|\Psi_{i}\rangle_{AE}|\Psi_{i}\rangle_{EB}$, where $|\Psi_{i}\rangle$ are Bell states:
\begin{eqnarray*}
|\Phi^{\pm}\rangle=|00\rangle\pm|11\rangle
\end{eqnarray*}
\begin{eqnarray*}
|\psi^{\pm}\rangle=|01\rangle\pm|10\rangle
\end{eqnarray*}
then, she can take all of keys that Alice sends to Bob. Here, we consider a simple case that the rotating angle by
Alice and Bob for the security of QKD protocol is equal to $\theta=\frac{\pi}{4}$, as a choice by ZLG protocol
(with a simple change in our approach, it can be applied to any arbitrary rotation angle $\theta$; although, some
authors consider Hadamard transformation for their operation in higher dimensions \cite{KBB} and more than two
parties \cite{BK}).

Eve's strategy can be described as follows:

1) Alice uses the particle $A$ of the entangled pairs and $\gamma_{1}$ in state $|\psi\rangle$ $(|0\rangle$ or
$|1\rangle)$ to do a \textsc{cnot} operation ($A$ is the controller and $\gamma_{1}$ is the target) and the three
particles will be in a GHZ (eqs. \ref{3}, \ref{4}) state. Then, she sends $\gamma_{1}$ to Bob.

2) Eve intercepts particle $\gamma_{1}$ and does a measurement in the $z$ bases $(|0\rangle, |1\rangle)$ on it.
Eve knows that Alice and Bob entangled states $|\Phi^{+}\rangle_{AB}$ reduce to $|00\rangle_{AB}$ or $|11\rangle_{AB}$.

3) Eve sends a particle $\gamma_{1}$ to Bob. Bob uses his corresponding particle $B$ to do a \textsc{cnot} operation on
$\gamma_{1}$ ($B$ is the controller and $\gamma_{1}$ is the target). Bob measures $\gamma_{1}$ and will get the
classical bit corresponding to the state $|\psi\rangle$. Alice and Bob can not detect Eve's operation in this step.

4) Alice and Bob apply the operation of $R(\frac{\pi}{4})$ on the their particles states. Alice chooses another particle
$\gamma_{2}$ $(|0\rangle$ or $|1\rangle)$ and does a \textsc{cnot} operation on it ($A$ is the controller and
$\gamma_{2}$ is the target), and sends $\gamma_{2}$ to Bob.

5) Eve intercepts and keeps particle $\gamma_{2}$ to herself. Thus, Eve has a share on the Bell state between herself
and Alice.

6) Eve prepares a maximally entangled state $|\Phi^{+}\rangle_{1,2}$ and sends particle $2$ to Bob. At this
stage, the common state among them is:
\begin{eqnarray*}
|\Psi_{1}\rangle=(|00\rangle-|11\rangle)_{A\gamma_{2}}(|00\rangle+|11\rangle)_{1,2}(|0\rangle
-|1\rangle)_{B}
\end{eqnarray*}
for $|\Phi^{+}\rangle_{AB}$ reduces to $|11\rangle_{AB}$ and $|\gamma_{2}\rangle=|1\rangle$, or
\begin{eqnarray*}
|\Psi_{2}\rangle=(|01\rangle-|10\rangle)_{A\gamma_{2}}(|00\rangle+|11\rangle)_{1,2}(|0\rangle
-|1\rangle)_{B}
\end{eqnarray*}
for $|\Phi^{+}\rangle_{AB}$ reduces to $|00\rangle_{AB}$ and $|\gamma_{2}\rangle=|0\rangle$, or
\begin{eqnarray*}
|\Psi_{3}\rangle=(|00\rangle+|11\rangle)_{A\gamma_{2}}(|00\rangle+|11\rangle)_{1,2}(|0\rangle
+|1\rangle)_{B}
\end{eqnarray*}
for $|\Phi^{+}\rangle_{AB}$ reduces to $|00\rangle_{AB}$ and $|\gamma_{2}\rangle=|1\rangle$, or
\begin{eqnarray*}
|\Psi_{4}\rangle=(|01\rangle+|10\rangle)_{A\gamma_{2}}(|00\rangle+|11\rangle)_{1,2}(|0\rangle
+|1\rangle)_{B}
\end{eqnarray*}
for $|\Phi^{+}\rangle_{AB}$ reduces to $|11\rangle_{AB}$ and $|\gamma_{2}\rangle=|0\rangle$

7) Bob takes particle $2$ and uses particles $B$ and $2$ to do a
\textsc{cnot} operation ($B$ is the controller and $2$ is the
target). Here, common state is:
\begin{widetext}
\begin{eqnarray*}
|\Psi_{1}\rangle=(|00\rangle-|11\rangle)_{A\gamma_{2}}\{(|00\rangle-|11\rangle)_{1B}
|0\rangle_{2} + (|01\rangle-|10\rangle)_{1B}|1\rangle_{2}\}
\end{eqnarray*}
or
\begin{eqnarray*}
|\Psi_{2}\rangle=(|01\rangle-|10\rangle)_{A\gamma_{2}}\{(|00\rangle-|11\rangle)_{1B}
|0\rangle_{2} + (|01\rangle-|10\rangle)_{1B}|1\rangle_{2}\}
\end{eqnarray*}
or
\begin{eqnarray*}
|\Psi_{3}\rangle=(|00\rangle+|11\rangle)_{A\gamma_{2}}\{(|00\rangle+|11\rangle)_{1B}
|0\rangle_{2} + (|01\rangle+|10\rangle)_{1B}|1\rangle_{2}\}
\end{eqnarray*}
or
\begin{eqnarray*}
|\Psi_{4}\rangle=(|01\rangle+|10\rangle)_{A\gamma_{2}}\{(|00\rangle+|11\rangle)_{1B}
|0\rangle_{2} + (|01\rangle+|10\rangle)_{1B}|1\rangle_{2}\}
\end{eqnarray*}
\end{widetext}

8) Bob does a measurement on the particle $2$ in the $z$
bases $(|0\rangle, |1\rangle)$. With the probability of one-half,
the above states reduce to:
\begin{eqnarray*}
|\Psi_{1}\rangle=(|00\rangle-|11\rangle)_{A\gamma_{2}}(|00\rangle-|11\rangle)_{1B}
\end{eqnarray*}
or
\begin{eqnarray*}
|\Psi_{2}\rangle=(|01\rangle-|10\rangle)_{A\gamma_{2}}(|01\rangle-|10\rangle)_{1B}
\end{eqnarray*}
or
\begin{eqnarray*}
|\Psi_{3}\rangle=(|00\rangle+|11\rangle)_{A\gamma_{2}}(|00\rangle+|11\rangle)_{1B}
\end{eqnarray*}
or
\begin{eqnarray*}
|\Psi_{4}\rangle=(|01\rangle+|10\rangle)_{A\gamma_{2}}(|01\rangle+|10\rangle)_{1B}
\end{eqnarray*}
At this stage, Eve shares two similar Bell states between herself
on the one hand, and Alice and Bob on the other hand. Thus, for the next rounds of the protocol Eve can take all of
keys that encrypted by Alice to $|\psi\rangle_{A\gamma_{2}}$ state and she sends those keys to Bob by encryption to
$|\psi\rangle_{1B}$, without being detected by them.

9) For the next rounds of the ZLG protocol, in the cases of $|\Psi_{1}\rangle$, Eve gets the results $|\gamma_{n}+1\rangle$
($|\gamma_{n}\rangle$) for $n$ odd (even) sequences of Alice and Bob results and viceversa for $|\Psi_{4}\rangle$.
Hence, in the case of $|\Psi_{2}\rangle$ $(|\Psi_{3}\rangle)$, Eve gets the results of  $|\gamma_{n}+1\rangle$
($|\gamma_{n}\rangle$) for all of $n$ sequences of Alice and Bob results. However, when Alice and Bob compare a
subsequence of the data bits publicly to detect eavesdropping, Eve leaks useful information for correcting her results.
More specifically, for any odd (even) numbers of data bit that is
announced by Alice and Bob, Eve can determine which of the four possible $|\Psi_{i}\rangle$ states is common among them.
By this means, Eve can obtain all of numbered data bits completely, without being detected by Alice and Bob.

10) Thus, with the probability $\frac{1}{2}$, Eve can get favorable results. In other cases (with the probability
$\frac{1}{2}$), the Alice and Bob can detect Eve's attack. In other word, in any two rounds of the experiment, Eve can
hope to get keys in one of them.

Recently, Gao \emph{et al.} \cite{Gao} (GGWZ) showed that ZLG protocol and its extensions are not secure. They presented
an eavesdropping strategy which allows Eve to obtain half of the data bits without being detected by the communicating
parties. Thus, they concluded that use of Hadamard gate or rotation by $\theta=\frac{\pi}{4}$, on each shared qubit
is not effective in prevent eavesdropping. This eavesdropping strategy is very simple and Alice and Bob can
detect the effect of Eve by a simple change of their protocol. As authors suggested their eavesdropping can be
removed by the choice of another rotation angle for $R(\theta)$ ($\theta\neq\frac{\pi}{4}$), by Alice and Bob. Hence,
the parties can consider other states such as:

\begin{eqnarray}
|\psi_{0}\rangle=\alpha|0\rangle+\beta|1\rangle, \hspace{.5cm}
|\psi_{1}\rangle=\beta|0\rangle-\alpha|1\rangle,\nonumber\\
\langle\psi_{i}|\psi_{j}\rangle=\delta_{i,j}
\hspace{.7cm}i,j=1,2
\end{eqnarray}
as the carrier state in the ZLG protocol, where $\alpha$ and $\beta$ are real variables with the properties:
$\alpha\neq\beta$, $\alpha,\beta\neq0$ and $\alpha^{2}+\beta^{2}=1$. Their protocol would be secure against of GGWZ
attack \cite{Gao} and Alice and Bob can use Hadamard or $R(\theta)=\frac{\pi}{4}$ as their operations.
If we follow the ZLG protocol with the above set of states, after Alice operation (by choosing $\psi_{0}$ as carrier),
we have:
\begin{eqnarray}
|\Psi\rangle=\sqrt{\frac{1}{2}}[|00\rangle|\psi_{0}\rangle+|11\rangle(2\alpha\beta|\psi_{0}\rangle+
(\beta^{2}-\alpha^{2})|\psi_{1}\rangle)]
\end{eqnarray}
This state has an interesting property: carrier states corresponding to states $|00\rangle$ and $|11\rangle$ are not
orthogonal to each other, and Eve can not perfectly entangle her state to those of Alice and Bob. Thus, for
every round of the protocol, Eve makes some errors on parties' measurement results. With attention to the aforementioned
strategy for the protection of the ZLG protocol against the GGWZ attack, in the following, we don't consider this eavesdropping
strategy for our proposed protocols which repair ZLG, KBB and Bk protocols.

Hence, it is not complicated to show that, in our approach to Eve's eavesdropping strategy, it is not important
whether Alice and Bob take an arbitrary rotation angle $\theta$ or use states of equation (6).
\emph{In both cases, Eve can still take an appropriate rotation angle and measurement bases that help her to do
eavesdropping.}

\section{Quantum key distribution for d-level systems with generalized Bell states }
Recently, Karimipour \text{\em{et al.}} \cite{KBB} extended ZLG protocol to $d$-level systems. Their protocol is based
on shared entanglement of a reusable Bell state. The security against some individual attacks is proved,
where the information gain of Eve is zero and the quantum bit error rate (QBER) introduced by her intervention is $(d-1)/d$. However, in this
section we show that Eve can attack by a special strategy to get, with probability $1/d$, all of keys without
being detected by Alice and Bob.\\
For convenience, we use the same notations as in \cite{KBB}. Let us first give a brief description of the QKD protocol
involved here. At the beginning, Alice and Bob share a generalized Bell state:
\begin{eqnarray}
|\psi_{00}\rangle=\frac{1}{\sqrt{d}}\sum_{j=0}^{d-1}|j,j\rangle_{a,b}
\end{eqnarray}
The qudit to be sent is denoted by $q$, which is encoded as a basis state $|q\rangle_{k}$. They defined controlled-right
and  controlled-left shift gates that play the role of controlled-\texttt{NOT} gate, and act as follows:
\begin{eqnarray}
R_{c}|i,j\rangle=|i,j+i\rangle,\hspace{.5cm}L_{c}|i,j\rangle=|i,j-i\rangle\hspace{.5cm} (mod \hspace{.2cm} d)
\end{eqnarray}
Alice performs a controlled- right shift on $|q_{1}\rangle_{k}$ and thus entangles this qudit to the previously
shared Bell state, producing the state:
\begin{eqnarray}\label{10}
|\Phi\rangle=\frac{1}{\sqrt{d}}\sum_{j=0}^{d-1}|j,j,q_{1}+j\rangle_{a,b,k} \label{1}
\end{eqnarray}
Then, she sends the qudit to Bob. At the destination, Bob performs a controlled-left shift on the qudit and disentangles it
from the Bell stae, hence revealing the value of $q$ with certainty. To protect this protocol against a specific kind of
attack, Alice and Bob proceed as follows. Before sending each of the qudits, Alice and Bob act on their shared Bell state
by the Hadamard gates $H$ and $H^{\ast}$, which are defined as
$H=\frac{1}{\sqrt{d}}\sum_{k,l=0}^{d-1}e^{\frac{i2\pi}{d} kl}|l\rangle\langle k|$, respectively. The key point is that a
Bell state $|\psi_{0,0}\rangle$ disentangled from the outside world is unchanged under this operation, while a state
entangled with outside is not:
\begin{eqnarray}
H\otimes H^{\ast}|\psi_{0,0}\rangle=|\psi_{0,0}\rangle.
\end{eqnarray}

Eve's strategy is similar to that in the previous section and can be described as follows:

1) Alice uses the particle $a$ of the entangled pairs and $q_{1}$ in state $|q_{1}\rangle$ to do a controlled- right
shift on it to produce the state (eq. \ref{10}) . Then, she sends $q_{1}$ to Bob.

2) Eve intercepts particle $q_{1}$ and does a measurement in the $z$ basis on it. Eve knows that Alice, Bob and particle
$q_{1}$ states reduce to $|j,j,q_{1}+j\rangle_{a,b,k}$ where $j=0,...d-1$.

3) Eve sends particle $q_{1}$ to Bob. Bob uses his corresponding particle $b$ to do a controlled-left shift on the
$q_{1}$. Bob measures $q_{1}$ and gets the classical dit corresponding to state the $|q_{1}\rangle$. At this stage,
Alice and Bob can not detect Eve's operation.

4) Alice and Bob apply the operations $H$ and $H^{\ast}$ on the their particle states.
\begin{eqnarray*}
H|j\rangle_{a}=\frac{1}{\sqrt{d}}\sum_{k,l=0}^{d-1}e^{\frac{i2\pi}{d} kl}|l\rangle\langle k |j\rangle_{a}
=\frac{1}{\sqrt{d}}\sum_{l=0}^{d-1}e^{\frac{i2\pi}{d} jl}|l\rangle_{a}\nonumber \\
H^{\ast}|j\rangle_{b}=\frac{1}{\sqrt{d}}\sum_{k,l=0}^{d-1}e^{-\frac{i2\pi}{d} kl}|l\rangle\langle k |j\rangle_{b}
=\frac{1}{\sqrt{d}}\sum_{l=0}^{d-1}e^{-\frac{i2\pi}{d} jl}|l\rangle_{b}
\end{eqnarray*}

5) Alice chooses another particle $q_{2}$ and performs a controlled- right shift on $|q_{2}\rangle_{k}$ and sends
$q_{2}$ to Bob.
\begin{eqnarray*}
\frac{1}{\sqrt{d}}\sum_{l=0}^{d-1}e^{-\frac{i2\pi}{d} jl}|l\rangle_{a}|q_{2}+l\rangle_{k}
\end{eqnarray*}

6) Eve intercepts and keeps particle $q_{2}$ to herself, thus, Eve has a share on the Bell state between herself and
Alice. Eve prepares a maximally entangled state $\frac{1}{\sqrt{d}}\sum_{j=0}^{d-1}|k,k\rangle_{1,2}$ and sends particle
$2$ to Bob.

7) Bob takes particle $2$ and uses particles $b$ and $2$ to do a controlled-left shift on it
\begin{eqnarray*}
\frac{1}{\sqrt{d}}\sum_{l,k=0}^{d-1}e^{-\frac{i2\pi}{d} jl}|l\rangle_{b}|k\rangle_{1}|k-l\rangle_{2}
\end{eqnarray*}
Bob does a measurement on the particle particle $2$ in the $z$ basis. With the probability of $1/d$,
we have $q_{2}=k-l$. Thus, at this stage, Eve shares two similar Bell states between herself
on the one hand and Alice and Bob on the other hand.
\begin{eqnarray*}
\frac{1}{\sqrt{d}}\sum_{l=0}^{d-1}e^{\frac{i2\pi}{d} jl}|l\rangle_{a}|q_{2}+l\rangle_{k}
\frac{1}{\sqrt{d}}\sum_{l=0}^{d-1}e^{-\frac{i2\pi}{d} jl}|l\rangle_{b}|q_{2}+l\rangle_{1}
\end{eqnarray*}

8) For the next rounds of KBB protocol, in the case of the above state,
Eve gets the results $|q_{n}-j\rangle$ ($|q_{n}-q_{2}\rangle$) for
$n$ odd (even) sequences of Alice and Bob results for fix amount
of $j=0,...d-1$ and $q_{2}=0,...,d-1$. However, when Alice and
Bob compare a subsequence of the data dits publicly to
detect eavesdropping, Eve leaks useful information for correcting
her results. More specifically, for any odd (even) numbers of data
dit that is announced by Alice and Bob, Eve can determine which of
the $d$ possible states is common among them. By this means, Eve
can obtain all of numbered data dits completely without being detected by
Alice and Bob.

9) Thus, with the probability $\frac{1}{d}$, Eve can get
favorable results. In other cases (with the probability
$\frac{d-1}{d}$), Alice and Bob can detect Eve's attack.

\section{NEW AND SECURE ENCRYPTION PROTOCOL }

In the preceding section, we showed that ZLG protocol for EPR
pairs acting as the quantum key to encode and decode the classical
cryptography is not secure against a specific eavesdropping
attack, the F-type attack. In this section we suggest three
modified protocols which are secure against F-type attack. Those protocols have some basic properties,
a) the previously shared Bell state is the main element of protocol, b) Alice and Bob use all Hilbert space dimension
communicating particle ($\gamma$) and all rounds of the protocol. c) parties don't have classical communication.

In what follows, we first suggest two QKD protocol basis on the reusable Bell state, but parties use classical
communication, in the tertiary protocol, we derive a efficient and secure QKD without
classical communication.

\subsection{ ZLG Protocol with Classical Communication}

In this protocol, we let Alice and Bob to have classical communications. The basic idea of our modification consists
of the following steps:

1) Alice and Bob want to share $N$ secure bits between themselves.
They have previously shared some quantity of the Bell pairs
$|\Phi^{+}\rangle_{AB}$. When the process begins, the two
particles rotate their state by the angle
$\theta=\frac{\pi}{4}$, respectively.

2) Alice randomly selects the value of a bit ($0$ or $1$) (totaly
$3N$ bits, $N$ for message-coding  and $2N$ for checking
sequences) and prepares a carrier particle $\gamma$ in the
corresponding states $|\psi\rangle$ ($|0\rangle$ or $|1\rangle$).
The classical bit and the state $|\psi\rangle$ are only known by
Alice herself.

3) Alice randomly selects one of the following manners to send the
qubit $|\psi\rangle$,

i) She uses the particle $A$ of the entangled pairs and $\gamma$
in state $|\psi\rangle$ to do a \textsc{cnot} operation ($A$ is
the controller and $\gamma$ is the target) and calls $N$ qbits of
it message-coding and $N$ qbits, checking.

ii) She operates with a $R(\frac{\pi}{4})$ on the particle $A$ of the
entangled pairs and takes the particle $\gamma$ in state
$|\psi\rangle$ to do a \textsc{cnot} operation ($A$ is the
controller and $\gamma$ is the target) and calls it the checking
($N$ qbit) sequence. Then, she sends $\gamma$ to Bob. Alice keeps
the arrangement of her choices to herself.

4) After Bob receives the particle $\gamma$ in the case of $ii$
(checking sequence) Alice informs Bob about her $R(\frac{\pi}{4})$
operation and Bob operates $R(\frac{\pi}{4})$ on the his qubit
$B$. Then, he uses his corresponding particle $B$ to do a
\textsc{cnot} operation ($B$ is the controller and $\gamma$ is the
target). Bob measures $\gamma$ and gets the classical bit
corresponding to the state $|\psi\rangle$. They keep measurement
results on $\gamma$ to themselves.

5) In the cases ($i$) Alice doesn't send any classical
information to Bob, and Bob uses his corresponding particle $B$ to
do a \textsc{cnot} operation ($B$ is the controller and $\gamma$
is the target). Then Bob measures $\gamma$ and gets the classical
bit corresponding to the state $|\psi\rangle$.

6) After the transmission of $3N$ bit sequence, Alice informs Bob
about her choices sequence for checking qubits ($2N$ qubits). They
compare their results to check eavesdropping. If the error rate
in this checking is below a certain threshold, then, they can take
message-coding qubits ($N$ qubits) as raw keys.

\subsection{ZLG Protocol with Direct Communication}
At this stage, we represent another protocol with the same result
as that of the previous approach. The first and second steps of
this protocol are similar to the previous one and we replace the
third and fourth steps as follows:

$3'$) Alice randomly selects one of the following manners,

i) She uses the particle $A$ of the entangled pairs and $\gamma$
in the state $|\psi\rangle$ as does a \textsc{cnot} operation ($A$
is the controller and $\gamma$ is the target) and calls it
message-coding ($N$ qbits) and checking ($N$ qbits).

ii) She doesn't do any operation on the $\gamma$ and calls it the
checking ($N$ qbits) sequence. Then, she sends $\gamma$ to Bob.
Alice keeps the arrangement of her choices to herself.

$4'$) After Bob receives the particle $\gamma$ in the case of $ii$
(checking sequence), Alice informs Bob not to do any operation on
$\gamma$. Thus,  Bob measure $\gamma$ and gets the classical bit
corresponding to the state $|\psi\rangle$. They keep the
measurement results on $\gamma$ to themselves.

The fifth and sixth steps are the same as in the previous
protocol.

\subsection{ZLG Protocol with the Use of Higher Dimensional Bell States}
In the Two previous QKD protocols, Alice and Bob restricted themselves to have two different types of states for security
and some classical communication for some rounds of the protocol. At this subsection, we suggest the new protocol that
doesn't use classical communication (as the original ZLG protocol). Every repairing of ZLG protocol must have three basic properties,

I) Parties don't have classical communication,

II) Uses previously shared reusable Bell state,

III) Uses orthogonal state as carrier.

With the analysis of our eavesdropping attack, the main defect of ZLG protocol is the reduction of Bell pairs to product
state by Eve's measurement on the carrier particle. Thus, we choose some entangled state which doesn't have this property.
For example, we consider higher dimensional systems for sharing Bell states that don't reduce to a product state
under Eve's measurement.

In what follows, we need to define the controlled gate $U_{c}$ between one qudit (as controller) and one qubit (as target).
Similar to KBB protocol \cite{KBB}, we consider the controlled $U_{c}$ that acts on the second qubit, conditioned on the
first qudit, and defined as:
\begin{eqnarray}\label{cnot}
U_{c}|i\rangle_{d}\otimes|j\rangle_{2}=|i\rangle_{d}\otimes U^{i}|j\rangle_{2}
\end{eqnarray}
In the above relation subscripts of the state vectors represent the dimension of Hilbert space corresponding to each
particle, and $U=\sigma_{x}$ for the spin half target particles. Thus, for $i$ even (odd) target state changes to $|j\rangle_{2}$
($|j+1\rangle_{2}$).

In our protocol, Alice and Bob have previously shared some quantity of the Bell pairs, serving
as the quantum key
\begin{eqnarray}
|\psi_{0,0}\rangle=\frac{1}{\sqrt{D}}\sum_{j=0}^{D-1}|j,j\rangle_{AB}
\end{eqnarray}
When the process begins, the two parties operate Hadamard gate on the their particle's state which were defined in the
sec. III. The state $|\psi_{0,0}\rangle$ does not change under a bilateral operation. Then Alice selects a value of a
bit ($0$ or $1$) and prepares a carrier particle $\gamma$ in the corresponding state $|q\rangle$
($|0\rangle$ or $|1\rangle$), randomly. The classical bit and the state $|q\rangle$ are only known to Alice
herself. Alice uses the particle $A$ of the entangled pairs and $\gamma$ in state $|q\rangle$ to do a
controlled gate $U_{c}$ operation and the three particles will be in state
\begin{eqnarray*}\label{14}
|\Psi\rangle=\frac{1}{\sqrt{D}}\left\{\sum_{j_{even}=0}|j,j,q\rangle_{AB\gamma}+
\sum_{j_{odd}=1}|j,j,q+1\rangle_{AB\gamma}\right\}\nonumber
\end{eqnarray*}
\begin{eqnarray}
\end{eqnarray}
Then she sends $\gamma$ to Bob. Bob uses his corresponding particle $B$ to do a controlled gate $U_{c}$ operation
on $\gamma$ again. Now the key particles $A$ and $B$ and the carrier particle $\gamma$ are in the
same state as the initial state:

\begin{eqnarray}
|\Psi'\rangle=|\psi_{0,0}\rangle_{AB}\otimes|q\rangle_{\gamma}
\end{eqnarray}
Bob measures $\gamma$ and gets the classical bit corresponding to the state $|q\rangle$.

At the intermediate stage of eq.(\ref{14}), if Eve intercepts particle $\gamma$ and does a measurement in the $z$
basis $(|0\rangle, |1\rangle)$ on it, $|\Psi_{0,0}\rangle_{AB}$ reduces to
$|\psi_{0,0}^{even}\rangle=\sum_{j_{even}=0}|j,j\rangle_{AB}$ or
$|\psi_{0,0}^{odd}\rangle=\sum_{j_{odd}=1}|j,j\rangle_{AB}$. Thus, Eve can't disentangle Alice and Bob states,
if Alice and Bob apply Hadanmard transformation on the their states (for example we consider $|\psi_{0,0}^{even}\rangle$),
we have:
\begin{eqnarray}
H\otimes
H^{\ast}|\psi_{0,0}^{even}\rangle=\sum_{l_{1},l_{2}}\Delta
(l_{i},d)|l_{1},l_{2}\rangle_{AB}
\end{eqnarray}
In the above equation $\Delta(l_{i},d)=\delta (l_{1}-l_{2})+\delta (l_{1}-l_{2}-d)+\delta
(l_{1}-l_{2}+d)$ and we restrict ourselves to $D=2d$ (even dimensional Hilbert space). Thus, the above state reduces to
$\sum_{l}|l\rangle_{A}[|l\rangle_{B}+|l+d\rangle_{B}+|l-d\rangle_{B}]
=\sum_{l}|l\rangle_{A}[|l\rangle_{B}+2|l+d\rangle_{B}]$. By noting the
above state, Eve can do any operation on the carrier state $\gamma$, for every cases. Alice and Bob have a chance
to detect Eve's operation in the next rounds of protocol. From the aforementioned reasons, we can deduce that the
F-attack does not apply to this approach, Eve can only entangle her state with Alice and Bob states and she wishes
that after her operation and parties operation, Alice and Bob can not detect her effects. This strategy is the same
as GGWZ attack which we have shown to be removable easily.

Generalizing this approach to the KBB protocol is simple, we consider carrier states in the $k$-dimensional Hilbert
space and shared Bell state in the $D$-dimensional Hilbert space. We divide the $D$-dimensional Hilbert space to $k$
sections, where any section has $d$ elements ($D=kd$), similar to KBB protocol. Alice performs a controlled-right
shift on $|r_{1}\rangle_{k}$ and thus entangles this qudit to the previously shared Bell state, producing the state:
\begin{widetext}
\begin{eqnarray}\label{1}
|\Phi\rangle=\frac{1}{\sqrt{D}}\left\{\sum_{j_{0}}|j_{0},j_{0}\rangle|r_{1}+j_{0}\rangle+...
+\sum_{j_{i}}|j_{i},j_{i}\rangle|r_{1}+j_{i}\rangle
+...+\sum_{j_{k-1}}|j_{k-1},j_{k-1}\rangle|r_{1}+j_{k-1}\rangle
\right\}
\end{eqnarray}
\end{widetext}
In the above relation, $|r_{1}+j_{i}\rangle$ calculated at $mod$ $k$ and summation on the $j_{i}$ represent one set
of $j$'s which run as $\{i, i+k,..., i+(d-1)k\}$ ($0\preceq i \preceq k-1$).
Alice then sends the qudit to Bob. At the destination, Bob performs a controlled-left shift on the qudit and
disentangles it from the Bell state, hence revealing the value of $q$ with certainty.
Similar to the usual manner, Alice and Bob act on their shared Bell state by the Hadamard gates $H$ and $H^{\ast}$.

At the intermediate stage of eq.(\ref{1}), If Eve intercepts particle $r$ in the intermediate cases and does a measurement
in the $z$ base on it, then, $|\Psi_{0,0}\rangle_{AB}$ reduce to
$|\psi_{0,0}^{i}\rangle=\sum_{j_{i}}|j_{i},j_{i}\rangle$ $0\preceq i \preceq k-1$. Thus, Eve can't disentangle Alice
and Bob states and F-attack doesn't apply to this approach again. Although, Alice and Bob can not detect Eve's
operation, but, after Hadamard transformation by parties, the shared state is transform to:

\begin{eqnarray}\label{}
|\Phi\rangle&=&\sum_{l}|l\rangle
\left\{|l-(k-1)d\rangle+...+|l-d\rangle+|l\rangle\right.\\\nonumber
&&\left.+|l+d\rangle+...+|l+(k-1)d\rangle \right\}
\\\nonumber
&=&\sum_{l}|l\rangle
\left\{|l\rangle+2|l+d\rangle+...+2|l+(k-1)d\rangle\right\}.
\end{eqnarray}


It can be shown that for every round of the protocol, Alice and Bob have a chance to detect Eve's operation.
Although, GGWZ \cite{Gao} attack still exists for this generalization, based on what we did in sec. III, we
can use other states as carriers, to get secured against GGWZ attacks.




\section{Non-security of quantum secret sharing based on reusable GHZ states  }
In a recent paper \cite{BK}, Bagherinzhad and Karimipour (BK) proposed a quantum secret sharing protocol based on
reusable GHZ states. The security against both intercept-resend strategy and entangle-ancilla strategy was proved.
However, here we show that with the help of our intercept-resend strategy, Eve can obtain all of the data bits with
probability $\frac{1}{2}$ without being detected by the communication parties. For convenience, we use the same notations
as in Ref.\cite{BK}.

\begin{figure}
\centering
\includegraphics[height=11.55cm,width=8cm]{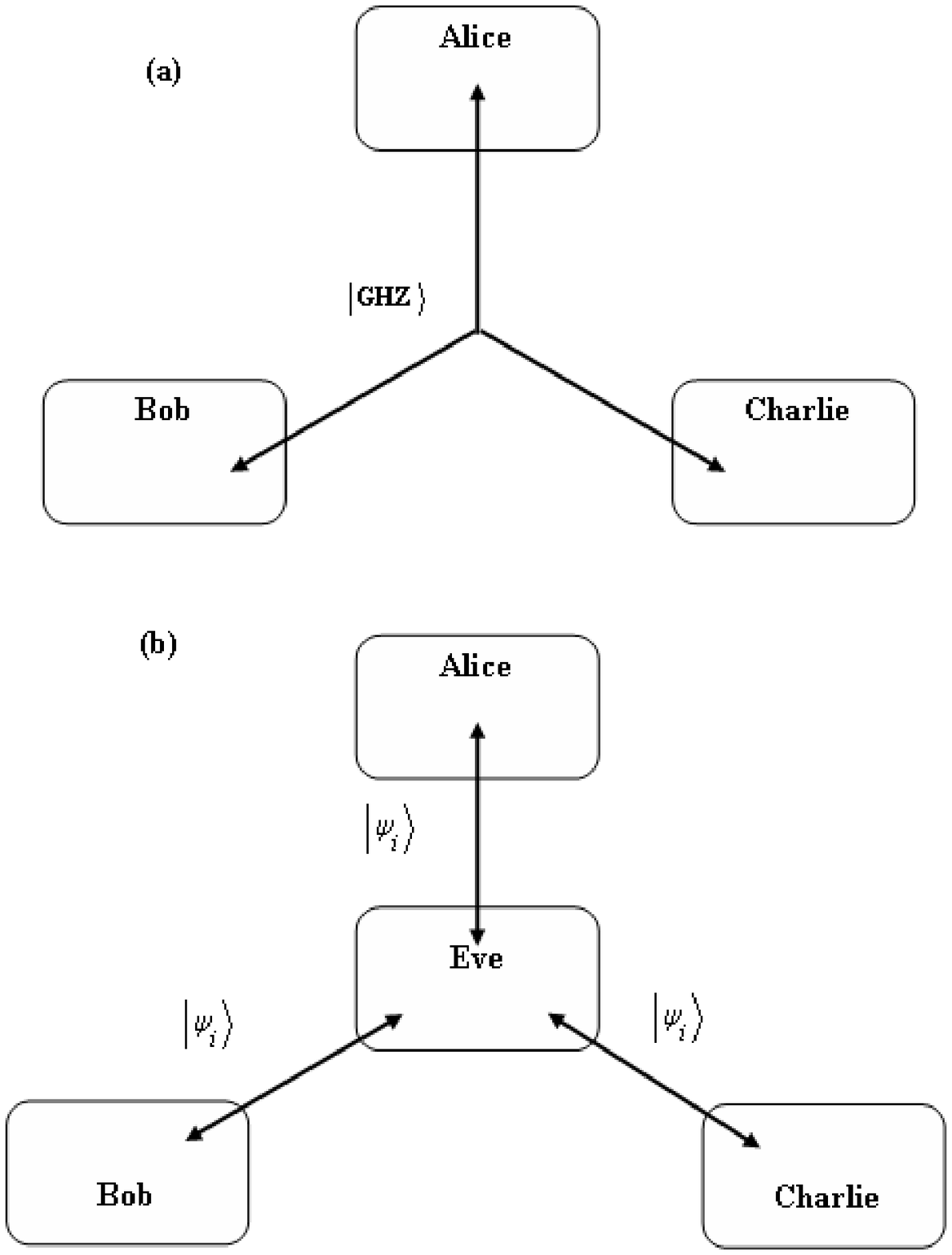}
\caption{A F-type attack on BK protocol for quantum encryption of
keys. (a) Alice, Bob ad Charlie previously shared entangled state
$|GHZ\rangle_{abc}$ among themselves. (b) Eve can transform
previously shared entangled state $|GHZ\rangle_{abc}$ to three
Bell states of
$|\Psi_{i}\rangle_{ae_{1}}|\Psi_{i}\rangle_{be_{2}}|\Psi_{i}\rangle_{ce_{3}}$,
among herself and parties, then, she can take all of keys that
Alice sends to Bob and Charlie.} \label{GHZ}
\end{figure}

The eavesdropping strategy is illustrated in Fig.(\ref{GHZ}). According to Fig.(\ref{GHZ}), if Eve can transform the
previously shared entangled state $|GHZ\rangle_{abc}$ among Alice, Bob ad Charlie to
$|\Psi_{i}\rangle_{ae_{1}}|\Psi_{i}\rangle_{be_{2}}|\Psi_{i}\rangle_{ce_{3}}$, then, she can take all of keys
that Alice sends to Bob and Charlie, where $|\Psi_{i}\rangle$ are Bell states. Now let us give an explicit description
of Eve's strategy.

1) In the beginning, Alice, Bob and Charlie have a perviously shared GHZ state which we denote by
\begin{eqnarray}
|G\rangle_{abc}:=\frac{1}{\sqrt{2}}(|000\rangle+|111\rangle)_{abc}
\end{eqnarray}
 and the data bits Alice wants to distribute to Bob and Charlie can be represented by
$q_{1},q_{2},q_{3},...,q_{n}$.

2) Alice prepares the state $|qq\rangle_{12}$ (corresponding to classical bit $q_{1}$) and entangles this state to the
already present GHZ state by performing \textsc{cnot} gates $C_{a1}C_{a2}$ (as defined at the KB protocol \cite{BK}) on
\begin{eqnarray}
|G\rangle_{abc}|qq\rangle_{12}=\frac{1}{\sqrt{2}}(|000\rangle+|111\rangle)_{abc}|q,q\rangle_{12}
\end{eqnarray}
to produce the state
\begin{eqnarray}
|\Phi^{odd}\rangle=\frac{1}{\sqrt{2}}(|000\rangle_{abc}|q,q\rangle_{12}+|111\rangle_{abc}|q+1,q+1\rangle_{12})\nonumber\\
\end{eqnarray}
and she sends the coded state to Bob and Charlie.

3) Eve intercepts first to send qubits and performs a measurement on them in the $z$ bases $(|0\rangle,|1\rangle)$.
Thus, the carrier state reduces to
\begin{eqnarray}
|\Phi^{odd}\rangle_{1}=|000\rangle_{abc}|q,q\rangle_{12}
\end{eqnarray}
or
\begin{eqnarray}
|\Phi^{odd}\rangle_{2}=|111\rangle_{abc}|q+1,q+1\rangle_{12}
\end{eqnarray}
Then, she resends the encoded states to Bob and Charlie. Bob and Charlie act on this state by the operators $C_{b1}$ and
$C_{c2}$ and extract the state $|q,q\rangle_{12}$. At this moment, parties can not detect Eve's operation.
In what follows, we only consider the $|\Phi^{odd}\rangle_{1}$ state (a similar approach can be
consider for the $|\Phi^{odd}\rangle_{2}$ state).

4) Alice, Bob and Charlie apply Hadamard operation $H$ on the their particle states. Then, for the even bits, which
are encoded as states $|\overline{q}\rangle$, i.e., $(|\overline{0}\rangle=(1/\sqrt{2})(|00\rangle+|11\rangle$ and
$|\overline{1}\rangle=(1/\sqrt{2})(|01\rangle+|10\rangle)$, Alice entangles this state to the shared state by
performing only one single \textsc{cnot} gate $C_{a1}$ on the $|\overline{q}\rangle$ to produce the state:
\begin{eqnarray*}
\frac{1}{\sqrt{2}}(|0\rangle_{a}|\overline{q}\rangle+|1\rangle|_{a}\overline{1+q}\rangle)\frac{1}{\sqrt{2}}(|0\rangle+|1\rangle)_{b}\frac{1}{\sqrt{2}}(|0\rangle+|1\rangle)_{c}
\end{eqnarray*}
Alice sends the coded state to Bob and Charlie.

5) Eve intercepts and keeps sending qubits to herself. Thus, she has shared the
$|0\rangle_{a}|\overline{q}\rangle+|1\rangle|\overline{1+q}\rangle_{a}$ state between herself and Alice.

6) Eve prepares two maximally entangled state $(|00\rangle+|11\rangle)_{34}(|00\rangle+|11\rangle)_{56}$ and sends
qubits 4 and 6 to Bob and Charlie respectively. At this stage, the common state between them is:
\begin{eqnarray*}
&&\frac{1}{\sqrt{2}}(|0\rangle_{a}|\overline{q}\rangle+|1\rangle|_{a}\overline{1+q}\rangle)\\
&&\times\frac{1}{2}(|0\rangle+|1\rangle)_{b}(|00\rangle+|11\rangle)_{34}
\frac{1}{2}(|0\rangle+|1\rangle)_{c}(|00\rangle+|11\rangle)_{56}
\end{eqnarray*}

7) Bob and Charlie take qubits 4 and 6 and act on the these states by operations $C_{b4}$ and $C_{c6}$. The common
state can be writhen as:
\begin{eqnarray*}
&&\frac{1}{\sqrt{2}}(|0\rangle_{a}|\overline{q}\rangle+|1\rangle_{a}|\overline{1+q}\rangle)\\
&&\times\frac{1}{2}\{(|00\rangle+|11\rangle)_{b3}|0\rangle_{4}+(|01\rangle+|10\rangle)_{b3}|1\rangle_{4}\}\\
&&\times\frac{1}{2}\{(|00\rangle+|11\rangle)_{c5}|0\rangle_{6}+(|01\rangle+|10\rangle)_{c5}|1\rangle_{6}\}
\end{eqnarray*}

8) Bob and Charlie do a measurement on the qubits 4 and 6 in the $z$ basis $(|0\rangle,|1\rangle)$, and identify
these types of keys together. With the probability of one-half, the above state reduces to:

for $q=0$
\begin{eqnarray*}
(|0\rangle_{a}|\overline{0}\rangle+|1\rangle_{a}|\overline{1}\rangle)
(|00\rangle+|11\rangle)_{b3}(|00\rangle+|11\rangle)_{c5}\\
(|0\rangle_{a}|\overline{0}\rangle+|1\rangle_{a}|\overline{1}\rangle)
(|01\rangle+|10\rangle)_{b3}(|01\rangle+|10\rangle)_{c5}
\end{eqnarray*}

for $q=1$
\begin{eqnarray*}
(|0\rangle_{a}|\overline{1}\rangle+|1\rangle_{a}|\overline{0}\rangle)
(|00\rangle+|11\rangle)_{b3}(|01\rangle+|10\rangle)_{c5}\\
(|0\rangle_{a}|\overline{1}\rangle+|1\rangle_{a}|\overline{0}\rangle)
(|01\rangle+|10\rangle)_{b3}(|00\rangle+|11\rangle)_{c5}
\end{eqnarray*}
Thus, Eve separately shares her qubit in a maximally entangled states with Alice's, Bob's and Charlie's.
Before going to the next round of QKD, we would like attention following relation for the above states for
$q=0,1$, we have:
\begin{eqnarray*}
H\otimes H\otimes H(|0\rangle|\overline{0}\rangle+|1\rangle|\overline{1}\rangle)=|000\rangle+|111\rangle\\
H\otimes H\otimes
H(|0\rangle|\overline{1}\rangle+|1\rangle|\overline{0}\rangle)=|000\rangle-|111\rangle
\end{eqnarray*}
and
\begin{eqnarray*}
H\otimes H(|0\rangle|0\rangle+|1\rangle|1\rangle)=(|00\rangle+|11\rangle)\\
H\otimes H(|0\rangle|1\rangle+|1\rangle|0\rangle)=(|00\rangle-|11\rangle)
\end{eqnarray*}
Thus, for odd sequence bits $q_{3},q_{5},...$ Eve gets the same bits that Alice sent (for each cases $q=0,1$).
and for even sequence bits $q_{4},q_{6},...$ Eve gets the bits $q_{4},q_{6},...$ ($q_{4}+1,q_{6}+1,...$ )
for $q=0 (q=1)$. On the other hand, Bob and Charlie get precisely the same sequence of bits that Alice sent for all
rounds of the QKD. Although, after parties' communication in the general channel for test of security, Eve can recognize
appropriate keys. This approach has an interesting property that Eve can get all of the secret keys with probability
one half (the same as in the case of ZLG protocol). In other words, the probability doesn't increase with going from two to
three parties.

In what follows, we would like to repair the BK protocol by using higher dimensional GHZ states. The new protocol that
we suggest is similar to our previous approach to the ZLG protocol. We consider a generalization of the GHZ state to a
$D$ dimensional system which was previously shared among the parties. This is the state define as:
\begin{eqnarray}
|\Psi^{odd}\rangle=\frac{1}{\sqrt{D}}\sum_{j=0}^{D-1}|j,j,j\rangle_{abc}
\end{eqnarray}
Alice wants to send the sequence of classical bits $q_{1},q_{2},...$ $(q_{i}=0,1)$ to Bob and Charlie. She prepares
the state $|q,q\rangle_{12}$ corresponding to $q_{1}$ and operates with controlled gates $U_{c}(a1)$ and $U_{c}(a2)$
(previously defined at eq. (\ref{cnot})) on them, to the get the state:
\begin{eqnarray}\label{25}
|\Psi^{odd}\rangle=\frac{1}{\sqrt{D}}\left\{\sum_{j_{even}=0}|j,j,j\rangle_{abc}|q,q\rangle_{12}\right.\nonumber\\
\left.+ \sum_{j_{odd}=1}|j,j,j\rangle_{abc}|q+1,q+1\rangle_{12}\right\}\nonumber\\
\end{eqnarray}
and she sends a coded state to Bob and Charlie. At the destination, Bob and Charlie act on this state by the controlled gates
$U_{c}(b1)$ and $U_{c}(c2)$ and extract the state $|q,q\rangle_{12}$ (for further convenience, we restrict our selves
to particles with even dimension $D=2d$). Hence, for the even bits which are encoded as state
$|\overline{q}\rangle$ (similar to BK protocol), parties apply Hadamard transformation to $D$-dimensional GHZ state
which we denote by
\begin{eqnarray*}
&&H \otimes H \otimes H|\Psi^{odd}\rangle
\\\nonumber
&&=|\Psi^{even}\rangle=\frac{1}{D}\sum_{l_{1},l_{2},l_{3}=0}^{D-1}
\Delta(l_{i},D)|l_{1},l_{2},l_{3}\rangle_{abc}
\end{eqnarray*}
In the above equation $\Delta(l_{i},D)=\delta (\sum_{i=1}^{3}l_{i})+\delta (\sum_{i=1}^{3}l_{i}-D)+\delta
(\sum_{i=1}^{3}l_{i}-2D)$. Thus, Alice entangles the state $|\overline{q}\rangle$ to the shared state
$|\Psi^{even}\rangle$ by performing only one single controlled gate $U_{c}(a1)$ to produce:

\begin{eqnarray}
&&\frac{1}{D} \left\{ \sum_{l^{even}_{1},l_{2},l_{3}=0}^{D-1}
\Delta(l_{i},D)|l_{1},l_{2},l_{3}\rangle_{abc}|\overline{q}\rangle\right.
\\\nonumber
&&\left.+\sum_{l^{odd}_{1},l_{2},l_{3}=0}^{D-1}\Delta(l_{i},D)|l_{1},l_{2},l_{3}\rangle_{abc}
|\overline{q+1}\rangle \right\}
\end{eqnarray}

At the destination, Bob and Charlie act on this state by the
operators $U_{c}(b1)$ and $U_{c}(c2)$ and extract
$|\overline{q}\rangle_{12}$. noting that $H^{2}\neq 1$ for $D>2$, Alice and Bob must operate $H^{\ast}$
in the next round of the protocol.

If in the intermediate step of eq. (\ref{25}), Eve intercepts the first sending qubits and performs a measurement on them,
then, the GHZ state reduces to:
\begin{eqnarray}
|\Phi_{even}\rangle=\sqrt{\frac{2}{D}}\sum_{j_{even}=0}|j,j,j\rangle_{abc}
\end{eqnarray}
or
\begin{eqnarray}
|\Phi_{odd}\rangle=\sqrt{\frac{2}{D} }\sum_{j_{odd}=1}|j,j,j\rangle_{abc}
\end{eqnarray}
If Alice, Bob and Charlie apply Hadamard gate on the their particle states, for example $|\phi_{even}\rangle$, we have:
\begin{eqnarray*}
H^{\ast} \otimes H^{\ast} \otimes H^{\ast}|\psi\rangle
=\sum_{h_{1},h_{2},h_{3}=0}^{D-1} \Delta
(h_{i},d)|h_{1},h_{2},h_{3}\rangle_{abc}
\end{eqnarray*}

In the above equation $\Delta(h_{i},d)=\delta (\sum_{i=1}^{3}h_{i})+\delta (\sum_{i=1}^{3}h_{i}-d)+\delta
(\sum_{i=1}^{3}h_{i}-2d)+\delta (\sum_{i=1}^{3}h_{i}-3d)+\delta (\sum_{i=1}^{3}h_{i}-4d)
+\delta (\sum_{i=1}^{3}h_{i}-5d)$. Thus, Eve can't disentangled Alice, Bob and Charlie states and we deduce that
F-attack doesn't apply to this approach It can be shown that for every round of the protocol, parties have a chance
to detect Eve's operation. Eve can only entangle her state with Alice and Bob states and she wishes that after her operation,
the parties, Alice, Bob and Charlie, can not detect her effects. This strategy is the same as the GGWZ \cite{Gao}
 attack, we showed to be removable easily.

One of interesting results of our extension is that our protocol can simply be extended to higher dimension QKD
between Alice , Bob and Charlie, by using quantum key encoding and decoding classical information.


\section{CONCLUSION}
In this Paper, we review ZLG, KBB and BK protocols for QKD via quantum encryption and show that these protocols are
not secure. Thus, Eve has always a chance to be informed of her parties' keys without being detected by them.
That is, we have shown that the security conditions are a necessary condition but not a sufficient one, specially in the
QKD with orthogonal states.

For the sake of completeness we briefly described the behavior of the presented schemes on efficiency of the transmission.
We consider the definition give by Cabello \cite{Cab}. The efficiency of a QKD protocol, $\varepsilon$, is defined as:
\begin{eqnarray}
\varepsilon=\frac{b_{s}}{q_{t}+b_{t}},
\end{eqnarray}
where $b_{s}$ is the expected number of secret bits received by Bob, $q_{t}$ is the number of qubits used in the
quantum channel, and $b_{t}$ is the number of bits used in the classical channel between Alice and Bob. Since in
our scheme no classical information is needed in the message-coding, we have $b_{t}=0$. This together with
$b_{s}=1$ and $q_{t}=1$ leads to $\varepsilon=1$.

However, the practical efficiency takes into account channel's transmittance as well \cite{Deg}. In our protocol
(ZLG protocol with higher dimensional Bell states) a qubit travels for a distance $L$, $L$ being the separation between
Alice and Bob. If $\tau$ is the transmittance of the qubit over a distance $L$, after traveling the distance $L$
and the operation of Bob, he takes measurement on the carrier particle. Thus, the practical efficiency can be evaluated as
$\varepsilon'=\varepsilon \tau^{3}=\tau^{3}$. By comparison, in BB84 it is $\varepsilon'= (1/6) \tau$,
because $b_{s}=0.5$, $q_{t}=1$ and $b_{t}=2$, and a qubit travels a distance $L$. This implies that
the presented scheme is more efficient than BB84, provided the transmittance of the channel is $\tau > \sqrt{1/6}$.
It is also possible to see that ours practical efficiency is more efficient than M. Lucamarini and S. Mancini protocol
\cite{Ma} if $\tau > \sqrt{1/2}$ and that the entanglement-based scheme \cite{Deg} contains a factor $\tau^{6}$,
making it lower than ours \cite{Comm}.

We would like to thank M. Golshani for useful comments and critical reading of the manuscript,
author is grateful to V. Karimipour for helpful discussions and also A. A. Shokri for his helps
(this work supported under project name: Gozaar).



\begin{references}
\bibitem {Comm} Although, in the \cite{Deg,Ma} authors claimed that their protocols have $\varepsilon=1$,
but with note to Cabello's definition for $\varepsilon$, it can be
shown that $\varepsilon=1/3,1/2$ respectively for their protocols,
here in the \cite{Deg} practical efficiency is equal to
$\varepsilon'= (1/3) \tau^{6}$.
\end{references}
\end{document}